\documentclass{aa}  

\usepackage{graphicx}
\usepackage{txfonts}
\usepackage{hyperref}
\def\detJ{\mathrm{det}J}

\def\tein{\theta_{\mathrm{Ein}}}
\def\trad{\theta_{\mathrm{rad}}}

\def\teinein{\theta_{\mathrm{Ein}}^*}
\def\betaein{\beta^*}

\def\toneobs{\theta_1^{\mathrm{obs}}}
\def\ttwoobs{\theta_2^{\mathrm{obs}}}

\def\moneobs{m_1^{\mathrm{obs}}}
\def\mtwoobs{m_2^{\mathrm{obs}}}

\def\asymm{\xi_{\mathrm{asymm}}}

\def\hyperpars{\boldsymbol{\eta}}

\def\nsource{N_{\mathrm{s}}}
\def\nbkg{n_{\mathrm{bkg}}}

\def\psilens{\boldsymbol{\psi}_\mathrm{g}}
\def\psisource{\boldsymbol{\psi}_\mathrm{s}}
\def\psisourcetwo{\boldsymbol{\psi}_{\mathrm{s},2}}
\def\psisourcens{\boldsymbol{\psi}_{\mathrm{s},\nsource}}
\def\psisourcenobeta{\boldsymbol{\psi}_\mathrm{s}^{(-\boldsymbol\beta)}}

\def\prlens{{\rm P}_\mathrm{g}}
\def\prsource{{\rm P}_\mathrm{s}}

\def\prsourceone{{\rm P}_{\mathrm{s},1}}
\def\prsourcetwo{{\rm P}_{\mathrm{s},2}}
\def\prsl{{\rm P}_\mathrm{{SL}}}
\def\prslone{{\rm P}_{\mathrm{SL},1}}
\def\prsltwo{{\rm P}_{\mathrm{SL},2}}

\def\pdet{{\rm P}_\mathrm{det}}
\def\pdetone{{\rm P}_{\mathrm{det},1}}
\def\pdettwo{{\rm P}_{\mathrm{det},2}}
\def\crosssect{\sigma_\mathrm{{SL}}}

\def\data{\mathbf{d}}
\def\datai{\mathbf{d}_i}

\def\dlens{\mathbf{d}_{\mathrm{g}}}
\def\dsource{\mathbf{d}_{\mathrm{s}}}

\def\mlim{m_{\mathrm{max}}}

\def\Sref#1{Section~\ref{#1}\xspace}
\def\Fref#1{Figure~\ref{#1}\xspace}
\def\Tref#1{Table~\ref{#1}\xspace}
\def\Eref#1{Equation~\ref{#1}\xspace}

\def\pr{{\rm P}}

\defcitealias{S+C21}{Paper~I}
\defcitealias{Son21}{Paper~II}
\setcitestyle{notesep={}}

\begin{document}

   \title{Statistical strong lensing. III. Inferences with complete samples of lenses}
   \titlerunning{Statistical strong lensing. III.}
   \authorrunning{Sonnenfeld}


   \author{Alessandro Sonnenfeld\inst{1}
          }

   \institute{Leiden Observatory, Leiden University, Niels Bohrweg 2, 2333 CA Leiden, the Netherlands\\
              \email{sonnenfeld@strw.leidenuniv.nl}
             }

   \date{}

 
  \abstract
    {
Existing samples of strong lenses have been assembled by giving priority to sample size, at the cost of having a complex selection function.
With the advent of the next generation of wide-field photometric surveys, however, it might become possible to identify subsets of the lens population with well-defined selection criteria, trading sample size for completeness.
}
   {
There are two main advantages of working with a complete sample of lenses.
First, it is possible to recover the properties of the general population of galaxies, of which strong lenses are a biased subset.
Second, the relative number of lenses and non-detections can be used to further constrain models of galaxy structure. 
This work illustrates how to carry out a statistical strong lensing analysis that takes advantage of these features.
} 
   {
I introduced a general formalism for the statistical analysis of a sample of strong lenses with known selection function, then tested it on simulated data.
The simulation consists of a population of $10^5$ galaxies with an axisymmetric power-law density profile, a population of background point sources, and a subset of $\sim10^3$ strong lenses, complete above an observational cut.
}
   {
The method allows to recover the distribution in Einstein radius and mass density slope of the galaxy population in an unbiased way.
The number of non-lenses helps to constrain the model when magnification data are not available.
}
   {
Complete samples of lenses are a powerful asset to turn precise strong lensing measurements into accurate statements on the properties of the general galaxy population.
}

   \maketitle
%

\section{Introduction}\label{sect:intro}

Strong gravitational lenses have been used extensively to study the structure and late ($z<1$) evolution of massive galaxies, both through the detailed study of a small number of systems \citep[see for example][]{Son++12,Bar++13,SLC15,Col++18,Sch++19}, and with the statistical combination of tens of objects \citep{Aug++10, Bol++12, Sch++14, ORF14, Son++15, O+A18}.
Strong lenses, however, are rare systems, and they usually require dedicated follow-up observations to confirm their nature and to obtain all of the necessary information for a successful analysis.
For these reasons, past lens finding campaigns have been carried out with the primary goal of maximising sample size while minimising costs, in terms of observational time.
This approach has made it possible, in the past decade, to carry out invaluable investigations of the structure of galaxies.
At the same time, however, it has made the selection function of the existing samples of lenses a highly non-trivial one.

Let us consider, for example, the Strong Lensing Legacy Survey \citep[SL2S,][]{Ruf++11,Gav++12,Son++13a,Son++13b}. SL2S lens candidates were first identified in ground-based imaging data, with the automatic lens-finding tool {\sc Ringfinder} \citep{Gav++14}, the performance of which was quantified on simulated lenses. Subsequently, lens candidates were visually inspected and ranked on the basis of their likelihood of being strong lenses, according to the judgement of strong lensing experts.
A few tens of the most promising lens candidates were followed-up with spectroscopic observations from $8m$-class telescopes and with high-resolution imaging data from the Hubble Space Telescope, with a higher priority given to lenses with larger photometric redshift.
Finally, the systems for which follow-up observations yielded measurements of the lens and source redshift were used for the science analyses.

The final sample was then the result of a combination of algorithms, human decisions, and the efficiency of obtaining source redshifts from spectroscopic observations.
The probability of a strong lens to be included in the SL2S sample is, at the very least, a function of Einstein radius, source surface brightness, image configuration, lens-source contrast, source redshift and source emission line strength.
Clearly, it is very difficult to obtain an accurate description of such a high-dimensional probability distribution.

The main problem of working with lens samples with an unknown selection function, is that it is difficult to generalise the measurements obtained with them to the population of galaxies from which they are drawn.
Strong lens samples are a biased subset of the general galaxy population, for two reasons.
First, galaxies that are more massive and more compact have a higher cross-section for strong lensing, and therefore are more likely to be strong lenses. 
Second, the probability of a lens to be included in a given survey depends on the efficiency of the survey at discovering it and obtaining all of the necessary data for its analysis.

In the next few years, upcoming wide-field surveys such as Euclid\footnote{\url{https://www.euclid-ec.org/}}, the LSST\footnote{\url{https://www.lsst.org/}} and the Chinese Space Station Telescope (CSST) will enable the discovery of order $10^5$ new strong lenses \citep{Col15}.
The first paper of this series \citep[][, hereafter Paper I]{S+C21}, however, showed how a sample of $10^3$ strong lenses can be sufficient to obtain precise and accurate measurements of several important parameters of the structure of galaxies, such as the stellar mass-to-light ratio and the dark matter density profile.
This means that, using the full set of $10^5$ Euclid-LSST-CSST lenses as starting point, there will be ample freedom in determining how such a sample is defined.
Depending on the efficiency of the lens finding algorithms that will be used, it might be possible to define subsamples of the lens population that are complete within a well-defined region in observational space.
The goal of this paper is to illustrate the advantages of working with such a complete subsample of lenses.

When the properties of the background source population are known, there are two benefits in having a complete lens sample.
The first is that it is possible to model the mapping between the general population of galaxies and that of the lens sample, thus removing the bias introduced by strong lensing.
The second is that the number of non-detections can be used along with the lensing data to further constrain models of galaxy structure.
In this work, I illustrate how to take advantage of these features in a statistical strong lensing inference.

Building on past work \citep{Son++15,Son++19}, I first introduce a general formalism for inferring the properties of the galaxy population from the analysis of a strong lensing survey with known selection function.
I then extend it to incorporate information from the number of non-lenses in the survey.
I test this method on a simulated complete set of lenses, generated under a series of simplifying assumptions.
Although additional work is needed to apply this formalism to real samples of lenses, this is the necessary first step towards enabling accurate inferences of galaxy properties with samples of strong lenses.

The structure of this work is the following.
In \Sref{sect:theory} I introduce the theoretical formalism on which this work is based.
In \Sref{sect:sims} I describe the simulations used to test the analysis method.
In \Sref{sect:model} I describe the model that I used to fit the simulated data.
In \Sref{sect:results} I show the results of the experiment.
I discuss the results in \Sref{sect:discuss} and draw conclusions in \Sref{sect:concl}.

The Python code used for the simulation and analysis of the lens sample can be found in a dedicated section of a GitHub repository\footnote{\url{https://github.com/astrosonnen/strong_lensing_tools}}.


\section{Theory}\label{sect:theory}

In this section I introduce the theoretical foundation for the analysis method that I use in this work.
In section \ref{ssec:defs} I define the quantities on which this method is based.
In section \ref{ssec:slonly} I explain how to infer the properties of a population of galaxies from the observation of a subset of strong lenses drawn from it.
In section \ref{ssec:nondet} I explain how to combine information from the strong lenses with non-detections.

\subsection{Definitions}\label{ssec:defs}

\subsubsection{Parameters and distributions}

Let us consider a strong lens survey that searches for lenses among a well-defined population of foreground galaxies and background sources, with a lens selection criterion $S$.
Let us assume that all of the properties of a galaxy that are relevant for determining its probability of being a lens can be summarised with a set of parameters $\psilens$.
This set of parameters will include the redshift of the galaxy and quantities describing its mass distribution. 
Let $\psisource$ be a set of parameters that are relevant for determining the probability of a background source to be strongly lensed.
These parameters will include the redshift of the source, and quantities describing its surface brightness distribution. 

Given a lens-source pair, $(\psilens,\psisource)$, let $\pdet(\psilens,\psisource|S)$ be the probability of the two objects to produce a lens that is detected by our survey, given the selection criterion $S$.
%
Finally, let $\prlens(\psilens)$ and $\prsource(\psisource)$ be the probability distribution of the foreground galaxy and background source population, respectively.
With these definitions, the (unnormalised) probability distribution of the population of strong lenses resulting from the survey is
\begin{equation}\label{eq:one}
\prsl(\psilens,\psisource|S) \propto \prlens(\psilens)\prsource(\psisource)\pdet(\psilens,\psisource|S).
\end{equation}

The main quantity of interest is the distribution of the properties of the lens galaxy population, $\prlens(\psilens)$.
Strong lensing data, however, can only directly constrain the left-hand side term, $\prsl(\psilens,\psisource|S)$.
This is, in a broad sense, an inverse problem.
As I show in this work, if the background source population $\prsource(\psisource)$ and the lens detection probability $\pdet(\psilens,\psisource|S)$ are known, it is possible to effectively invert \Eref{eq:one} and solve for $\prlens(\psilens)$.
\Tref{tab:pars} lists the definitions of all of the quantities introduced in this section, as well as those that are defined later in this paper.

\begin{table*}
\caption{Notations used in this work}
\label{tab:pars}
\begin{tabular}{cl}
\hline
\hline
Quantity & Description \\
\hline
$\psilens$ & Set of parameters describing a foreground galaxy. \\
$\psisource$ & Set of parameters describing a background source. \\
$\prlens(\psilens)$ & Probability distribution of the foreground galaxy population. \\
$\prsource(\psisource)$ & Probability distribution of the background source population. \\
$S$ & Selection criteria of the strong lens survey. \\
$\prsl(\psilens,\psisource|S)$ & Probability distribution of the strong lens population, given the survey selection criteria. \\
$\boldsymbol\beta$ & Angular position of the source. \\
$\psisourcenobeta$ & Set of background source parameters, excluding the angular position. \\
$\crosssect(\psilens,\psisourcenobeta|S)$ & Strong lensing cross-section of a galaxy with parameters $\psilens$, relative to a source with parameters $\psisourcenobeta$. \\
$\data$ & Array of observational data (of the entire survey, or relative to a generic system). \\
$\Omega$ & Volume of the data space defined by the selection criteria of the survey. \\
$\hyperpars$ & Set of parameters describing the population of foreground galaxies. \\
$\nsource$ & Number of strongly lensed sources associated to a foreground galaxy. \\
$\lambda$ & Expectation value of the number of strongly lensed sources associated to a galaxy. \\
$\nbkg$ & Projected number density of background sources. \\
$\pr(\psisource|{\rm Lens},S)$ & Probability distribution of the source parameters, given that the source is strongly lensed. \\
$\theta$ & Angular coordinate in the image plane of a strong lens. \\
$\alpha(\theta)$ & Deflection angle of the lens. \\
$\tein$ & Einstein radius of a lens, in angular units. \\
$\gamma$ & Slope of the density profile of a power-law lens. \\
$\theta_1,\theta_2$ & Coordinates of image 1 and 2 (the main image of a strongly lensed source and the counter image). \\
$\mathcal{N}(\mu,\sigma^2)$ & Normal distribution with mean $\mu$ and variance $\sigma^2$. \\
$m_{\mathrm{s}}$ & Source intrinsic (unlensed) magnitude. \\
$m_{\mathrm{s}}^*$ & Knee of the source magnitude distribution function. \\
$\alpha_{\mathrm{s}}$ & Faint-end slope of the source magnitude distribution function. \\
$\mu(\theta)$ & Lensing magnification at image position $\theta$. \\
$m_1,m_2$ & True magnitudes of images 1 and 2.\\
$\moneobs,\mtwoobs$ & Observed magnitudes of images 1 and 2.\\
$\epsilon_m$ & Observational uncertainty ($1\sigma$) on the image magnitudes. \\
$m_{\mathrm{max}}$ & Maximum allowed magnitude of image 2, for a system to be defined a strong lens. \\
$\mu_{\mathrm{Ein}}$ & Parameter describing the mean $\log{\tein}$ of the galaxy population. \\
$\sigma_{\mathrm{Ein}}$ & Parameter describing the scatter in $\log{\tein}$ of the galaxy population. \\
$\mu_\gamma$ & Parameter describing the mean $\gamma$ of the galaxy population. \\
$\sigma_\gamma$ & Parameter describing the scatter in $\gamma$ of the galaxy population. \\
$\asymm$ & Image configuration asymmetry, defined in \Eref{eq:asymm}. \\
$\teinein$ & Value of $\tein$ needed to reproduce the observed image positions, for a given $\gamma$. \\
$\betaein$ & Value of $\beta$ needed to reproduce the observed image positions, for a given $\gamma$. \\
$\detJ$ & Jacobian determinant of the variable change $(\log{\tein},\gamma) \rightarrow (\theta_1,\theta_2)$.\\
\end{tabular}
\end{table*}

\subsubsection{The lensing cross-section}

A very important parameter, that is necessary to evaluate the likelihood of strong lensing data, is the source position relative to the lens, $\boldsymbol\beta$.
This is a source parameter, therefore it is included in $\psisource$.
However, it is convenient to split $\psisource$ as follows:
\begin{equation}
\psisource = (\psisourcenobeta,\boldsymbol\beta),
\end{equation}
where $\psisourcenobeta$ includes all of the remaining source parameters.

Let us assume that the angular position in the sky does not correlate with any other source property.
Given a strong lens with parameters $(\psilens,\psisourcenobeta)$, the probability of the source position being $\boldsymbol\beta$ is proportional to
\begin{equation}
\prsl(\boldsymbol\beta|\psilens,\psisourcenobeta,S) \propto U_{\mathbb{R}^2}(\boldsymbol\beta) \pdet(\psilens,\psisourcenobeta,\boldsymbol\beta|S),
\end{equation}
where $U_{\mathbb{R}^2}(\boldsymbol\beta)$ is a uniform distribution over the sky.
The inverse of the normalisation constant of this distribution is 
\begin{equation}\label{eq:crosssect}
\crosssect(\psilens,\psisourcenobeta|S) = \int_{\mathbb{R}^2} d\boldsymbol\beta \pdet(\psilens,\psisourcenobeta,\boldsymbol\beta|S).
\end{equation}
I refer to this quantity as the strong lensing cross-section.

I point out that $\crosssect$ does not necessarily correspond to a well-defined region in the source plane, within which a source is always lensed. This is because the detection probability $\pdet$ can take any value between $0$ and $1$, meaning that certain lens-source pairs only give rise to detected lenses with some probability.

\subsubsection{Selection criteria}

In this work, I focus on a particular class of lens selection criteria, based on a cut in observational space.
Let $\data$ be an array that summarises the observational data that are relevant for the detection and modelling of a strong lens. 
This array will include, for instance, the observed lens and source redshift (formally, these are different from the true redshifts that are included in $\psilens$ and $\psisource$), the surface brightness distribution of the lens galaxy, the number of images of the strongly lensed source, their position and surface brightness distribution.
I then consider, as selection criterion, one that only selects lenses within a volume $\Omega$ of the multi-dimensional space spanned by the data array:
\begin{equation}
S = \data \in \Omega.
\end{equation}
An example of such a criterion is the following: only lens galaxies brighter than a minimum luminosity, within a given redshift range, with at least two strongly lensed images brighter than a given threshold are included in the sample.

Finally, I assume that it is possible to define a volume $\Omega$ in observational space and design a lens survey that retrieves all of the existing lenses in $\Omega$.
I refer to that as a complete lens sample over the volume $\Omega$.

\subsection{Inference with strong lenses alone}\label{ssec:slonly}

Let us assume that the galaxy distribution can be described with an analytical model, summarised by a set of population-level parameters $\hyperpars$:
\begin{equation}
\prlens(\psilens) = \prlens(\psilens|\hyperpars).
\end{equation}
The goal of this method is to infer the posterior probability distribution of $\hyperpars$ given the data from the ensemble of strong lenses, $\{\data\}$.
From Bayes' theorem, this is given by the product between the prior probability of $\hyperpars$ and the likelihood of observing the data given $\hyperpars$:
\begin{equation}
\pr(\hyperpars|\{\data\}) \propto \pr(\hyperpars)\pr(\{\data\}|\hyperpars).
\end{equation}
Measurements obtained on different lenses are independent from each other. Therefore, the likelihood term can be written as
\begin{equation}\label{eq:likeprod}
\pr(\{\data\}|\hyperpars) = \prod_i \pr(\datai|\hyperpars),
\end{equation}
where $\datai$ is the data array relative to the $i-$th lens.
The observed data depends only indirectly on the lens population parameters $\hyperpars$: only after specifying the parameters describing the lens and the source properties, $\psilens$ and $\psisource$, it is possible to evaluate the likelihood. Since these are not univocally set by $\hyperpars$, it is necessary to marginalise over all possible values taken by them. In other words, each factor in the product of \Eref{eq:likeprod} becomes
\begin{equation}\label{eq:likeint1}
\pr(\datai|\hyperpars) = \int d\psilens d\psisource \pr(\datai|\psilens,\psisource) \prsl(\psilens,\psisource|\hyperpars,S),
\end{equation}
where $\prsl(\psilens,\psisource|\hyperpars,S)$ is the normalised probability distribution of the strong lenses, given the galaxy population parameters $\hyperpars$ and the lens selection criterion $S$.
With these definitions, it is possible, in principle, to explore the parameter space of $\hyperpars$ and compute its posterior probability distribution.

\Eref{eq:likeint1} is very similar in form to Equation (30) of \citetalias{S+C21}. The main difference is that the formalism adopted in that case was aimed purely at inferring the properties of the lens galaxies. Here, instead, the goal is to recover the distribution of the general galaxy population, $\prlens(\psilens|\hyperpars)$, which enters the integrand function of \Eref{eq:likeint1} through the product of \Eref{eq:one}.

Evaluating integrals of the kind of \Eref{eq:likeint1} in an accurate way requires a good knowledge of the source population and the lens detection probability.
Having a complete lens set is not a strict requirement for this purpose, but, as I discuss later, simplifies the problem and guarantees robustness.

\subsection{Inference with lenses and non-detections}\label{ssec:nondet}

Section \ref{ssec:slonly} explained how to infer the properties of the galaxy population, summarised by the parameters $\hyperpars$, when using only data from detected strong lenses.
When considering also non-detections, we need to adopt a slightly different formalism.

First of all, the data array $\data$ is modified to include an additional observable: the number of strongly lensed sources, $\nsource$.
This is $\nsource=0$ for the non-detections and $\nsource=1$ (or larger, in the rare cases in which more than one source is strongly lensed by the same galaxy) for the lenses.
Second, the product of \Eref{eq:likeprod} is now calculated over the entire galaxy sample.
Third, only a fraction of the galaxies in the sample have a background source associated with them.
Given these considerations, it is useful to first expand the integrands of \Eref{eq:likeprod} in terms of the galaxy parameters only:
\begin{equation}\label{eq:likeintnosource}
\pr(\datai|\hyperpars) = \int d\psilens \pr(\datai|\psilens,S)\pr(\psilens|\hyperpars).
\end{equation}
In this case, the galaxy probability distribution is simply described by $\pr(\psilens|\hyperpars)$, which is the distribution of the general galaxy population.
This is because galaxies are included in the sample independently of whether they produce a detectable strong lens or not.

Let us consider the case in which a given galaxy is not a strong lens.
The only data available in this situation are the number of strongly lensed sources ($\nsource=0$) and observables related to the lens galaxy, which I indicate with $\dlens$.
The likelihood term $\pr(\datai|\psilens,S)$ can then be written as
\begin{equation}\label{eq:likenonlens}
\pr(\datai|\psilens,S)^{(\mathrm{Non-lens})} = \pr(\nsource=0|\psilens,S)\pr(\dlens|\psilens).
\end{equation}
The first factor on the right-hand side of the above equation is the likelihood of observing no strongly lensed sources, given the galaxy parameters and the selection criteria of the survey.
The probability of detecting $\nsource$ lensed sources is given by Poisson statistics:
\begin{equation}\label{eq:poiss}
\pr(\nsource|\psilens,S) = \frac{\lambda(\psilens,S)^{\nsource}e^{-\lambda(\psilens,S)}}{\nsource!},
\end{equation}
where $\lambda(\psilens,S)$ is the average number of strongly lensed sources that are expected for a galaxy with parameters $\psilens$ and a survey with selection criteria $S$.
In the limit $\lambda \ll 1$, usually satisfied in the galaxy-scale strong lensing regime, we have
\begin{align}
\pr(0|\psilens,\psisource,S) & \approx 1 - \lambda \nonumber \\
\pr(1|\psilens,\psisource,S) & \approx \lambda \\
\pr(\nsource > 1|\psilens,\psisource,S) & \approx 0 \nonumber ,
\end{align}
to first order in $\lambda$.
Assuming that the background sources are uniformly distributed in space with average projected number density $\nbkg$, then $\lambda(\psilens,S)$ is given by
\begin{equation}\label{eq:lambda}
\lambda(\psilens,S) = \nbkg \int d\psisourcenobeta \pr(\psisourcenobeta) \crosssect(\psilens,\psisourcenobeta|S),
\end{equation}
which is the product between $\nbkg$ and the strong lensing cross-section, averaged over the distribution of source parameters other than position, $\pr(\psisourcenobeta)$.

If a galaxy is a strong lens, then the data vector includes also observables related to the lensed source(s), $\dsource$.
In order to compute the likelihood term $\pr(\datai|\hyperpars)$, then, one must expand \Eref{eq:likeintnosource} by including also the source parameters. In the case of a single source, this becomes
\begin{align}\label{eq:lensdetint}
\pr(\datai|\hyperpars)^{(\mathrm{Lens})} = \int d\psilens d\psisource & \pr(\nsource|\psilens,S)\pr(\dlens|\psilens)\pr(\dsource|\psilens,\psisource)\times \\ \nonumber
& \pr(\psilens|\hyperpars)\pr(\psisource|{\rm Lens},\psilens,S).
\end{align}
In the integrand function, the probability of observing the data given the lens and source parameters and the selection criteria is written as the product of three factors.
The first factor is the probability of observing $\nsource$ lensed sources given the lens parameters and selection criteria, which is computed in the same way as in the non-detection case: first by calculating $\lambda$ via \Eref{eq:lambda}, then by taking the Poisson probability of \Eref{eq:poiss}.
The second factor in the integrand is the probability of observing galaxy-related data $\dlens$ given the parameters $\psilens$, while the third factor is the probability of observing the lensed image configuration $\dsource$ given the lens and source parameters.

Finally, the last factor in the integrand of \Eref{eq:lensdetint} is the probability distribution of the source parameters.
This factor must take into account the fact that the source is strongly lensed. Therefore, its unnormalised probability distribution is given by
\begin{equation}
\pr(\psisource|{\rm Lens},\psilens,S) \propto \prsource(\psisource)\pdet(\psilens,\psisource|S).
\end{equation}

There is a subtle difference between the source parameters $\psisource$ of \Eref{eq:lensdetint} and those entering the estimation of $\lambda$ in \Eref{eq:lambda}.
The former describe a well-defined background source that corresponds to the observed strongly lensed images.
The latter refer generically to the source population, and not to a specific source.
For this reason, the probability of observing $\nsource$ sources depends only on the lens parameters and not on those of its associated source.

On a related note, there need not be a strict correspondence between the number of detected sources and the number of sources that are taken into account in a lens model.
For example, if more than a single source is detected, one may decide to only explicitly model one of them, using \Eref{eq:lensdetint}, all of them, by expanding \Eref{eq:lensdetint} over new dimensions $\psisourcetwo,\ldots,\psisourcens$, or even none of them.
In this last case, the likelihood can be evaluated simply with \Eref{eq:likeintnosource}.


\section{Simulations}\label{sect:sims}

In this section I describe the simulations created to test the method introduced in \Sref{sect:theory}. The main purpose of these simulations is to show concrete examples of how to carry out an inference with a complete set of lenses.
In order to make these examples easy to follow, I created the simulations by making the following series of simplifying assumptions:
\begin{itemize}
\item All of the foreground galaxies have an axisymmetric mass distribution.
\item All of the background sources are point-like.
\item All of the foreground galaxies are at the same redshift, $z_g$.
\item All of the background sources are at the same redshift, $z_s$.
\end{itemize}
In section \ref{ssec:lenses} I describe the properties of the foreground galaxy population. In section \ref{ssec:sources} I describe the background source population. In section \ref{ssec:sldef} I describe the selection criterion of the resulting strong lens sample.


\subsection{Foreground galaxy population}\label{ssec:lenses}

Each foreground galaxy is described by an axisymmetric power-law mass distribution.
Let $\theta$ be the angular coordinate in the image plane along an axis with origin on the lens centre.
The lensing deflection angle of such a galaxy as a function of $\theta$ is
\begin{equation}
\alpha(\theta) = \tein \frac{\theta}{|\theta|}\left(\frac{|\theta|}{\tein}\right)^{2-\gamma},
\end{equation}
where $\tein$ is the Einstein radius and $\gamma$ is the density slope (the spherical deprojection of a power-law lens has a density profile $\rho(r) \propto r^{-\gamma}$).

As section 2.1 of \citetalias{S+C21} shows, the types of lensed image configurations that such a lens can produce depend on the value of $\gamma$. When $\gamma <2$, the lens has a non-degenerate radial caustic: sources located inside the caustic are strongly lensed into three images. Assuming without loss of generality that the source position $\beta$ is positive, the three images $(\theta_1,\theta_2,\theta_3)$ are located at $\theta_1 > \tein$, $-\tein < \theta_2 < -\trad$ and $-\trad < \theta_3 < 0$, where $\trad$ is the radius of the radial critical curve.

Lenses with $\gamma > 2$ always produce two multiple images, at $\theta_1 > \tein$ and $-\tein < \theta_2 < 0$. However, the magnification of image 2 goes to zero when the source position is large.
Finally, when $\gamma=2$ \citepalias[not shown in][]{S+C21}, sources located within $\beta < \tein$ are strongly lensed into two images.
For more details of the lensing properties of axisymmetric power-law lenses, I refer to section 2.1 of \citetalias{S+C21}.

 
I generated $10^5$ such galaxies, drawing $\log{\tein}$ and $\gamma$ from the following distributions:
\begin{align}
\log{\tein} & \sim \mathcal{N}(-0.3, 0.2^2) \\
\gamma & \sim \mathcal{N}(2.0, 0.2^2),
\end{align}
where $\mathcal{N}(\mu,\sigma^2)$ indicates a Gaussian with mean $\mu$ and variance $\sigma^2$.
The Einstein radius is in units of arcsec, and contains implicitly information on the lens and source redshift and on the cosmology.

\subsection{Background source population}\label{ssec:sources}

I obtained the distribution in apparent magnitude $m_s$ of the background sources as follows.
First, I considered the rest-frame $1500\AA$ luminosity function of galaxies at $1.5 < z < 2.5$ measured by \citet{Par++16}.
Then, I converted it to the $g$-band apparent magnitude assuming a fixed redshift of $z=1.5$, which is a typical value for a galaxy-scale strong lens, and a fixed spectral energy distribution, the `starb1` template from the Kinney-Calzetti Spectral Atlas of Galaxies \citep{Cal++94,Kin++96}.
The resulting apparent magnitude distribution is the following:
\begin{equation}\label{eq:msdist}
m_s \sim \left(10^{-0.4(m_s - m_s^*)}\right)^{\alpha_s+1} \exp{\left\{-10^{-0.4(m_s - m_s^*)}\right\}},
\end{equation}
with $m_s^*=24.47$ (also reerred to as the knee of the magnitude distribution) and $\alpha_s=-1.32$ (the faint-end slope).
For the sake of computational efficiency, I truncated the distribution outside of the range $m_s\in[23.0,28.0]$.
This cut does not affect the results: the number of galaxies brighter than $23$ in the distribution of \Eref{eq:msdist} is highly suppressed, and, because of the lens selection criterion that is introduced in section \ref{ssec:sldef}, the fraction of galaxies fainter than $28$ that results in detected strong lenses is also very small.

I generated, behind each lens, a random realisation of background source positions, drawn from a uniform distribution with average number density $\nbkg=100{\rm\,arcmin}^{-2}$.
Most of these sources are very faint, by the standard of current wide-field photometric surveys.
When considering only sources brighter than $26$, $\nbkg$ is a factor of four smaller and matches closely the number density of sources in the Hyper Suprime-Cam Subaru Strategic Program \citep[HSC survey][]{Aih++18} weak lensing shear catalogue \citep[approximately $25{\rm\,arcmin}^{-2}$, ][]{Man++18}.

\subsection{Selection criteria}\label{ssec:sldef}

For each lens, I defined a search region in the source plane to identify background sources that are multiply imaged. For lenses with $\gamma<2$, this search region was simply the area within the radial caustic. For lenses with $\gamma>2$, I set the limit to a very large value of $\beta$.
For each multiply-imaged source, I solved the lens equation,
\begin{equation}
\beta = \theta - \alpha(\theta),
\end{equation}
to find the image positions.
I assumed that only image 1 and 2 are observable, and discarded image 3, which, when present, is typically highly de-magnified.
I calculated the magnification $\mu$ of each image and used it to compute its magnitude:
\begin{align}
m_1 & = m_s - 2.5\log(|\mu(\theta_1)|) \\
m_2 & = m_s - 2.5\log(|\mu(\theta_2)|).
\end{align}
With $m_1$ and $m_2$ I indicate the true magnitudes of the two images.
Then, I added a random observational error of $\epsilon_m=0.1$~mag to $m_1$ and $m_2$ to generate observed magnitudes:
\begin{align}
m_1^{(\mathrm{obs})} & \sim \mathcal{N}(m_1,\epsilon_m^2) \\
m_2^{(\mathrm{obs})} & \sim \mathcal{N}(m_2,\epsilon_m^2).
\end{align}

Given these data, I defined as a strong lens any lens-source pair for which the observed magnitude of image 2 is brighter than a limiting magnitude $\mlim=26.6$.
The choice of this value is based on the lens search of \citet{Son++18} among data from the HSC survey: it corresponds to the $5\sigma$ detection limit for a point source in the $g$-band, which is the band used by \citet{Son++18} for the detection of lensed images.
The selection criterion is then
\begin{equation}\label{eq:maxmag}
S = \left\{m_2^{(\mathrm{obs})} < \mlim\right\}.
\end{equation}
The rationale for this choice is the following.
The process of finding lenses involves always a visual inspection step.
During this step, lensing experts estimate to what degree a lens candidate can be considered a real strong lens. Only lenses that are identified with high confidence are used for a scientific analysis. One very important element that experts look for in their classification is the presence of multiple images \citep[see for example][]{Son++20}. 
In all practical cases, image 1 is always brighter than image 2. Therefore, the condition of \Eref{eq:maxmag} ensures that both images are detected.

Finally, I assumed that the strong lens survey is complete above the limit imposed with \Eref{eq:maxmag}: all of the lenses with an observed magnitude brighter than image 2 are included in the sample; all of the remaining lenses are discarded and treated as non-detections.
Under these assumptions, the detection probability of a lens-source pair that produces two images is
\begin{equation}
\pdet(\tein,\gamma,m_s,\beta|S) = \frac{1}{2}\left[1 - {\mathrm{erf}}{\left(\frac{m_2 - \mlim}{\sqrt{2}\epsilon_m}\right)}\right].
\end{equation}
The generated sample consists of $867$ lenses, out of the initial $10^5$ galaxies. This is a typical value of the ratio between the number of lenses and the number of inspected galaxies in the HSC lens search \citep{Son++18}. 

\begin{figure*}
\includegraphics[width=\textwidth]{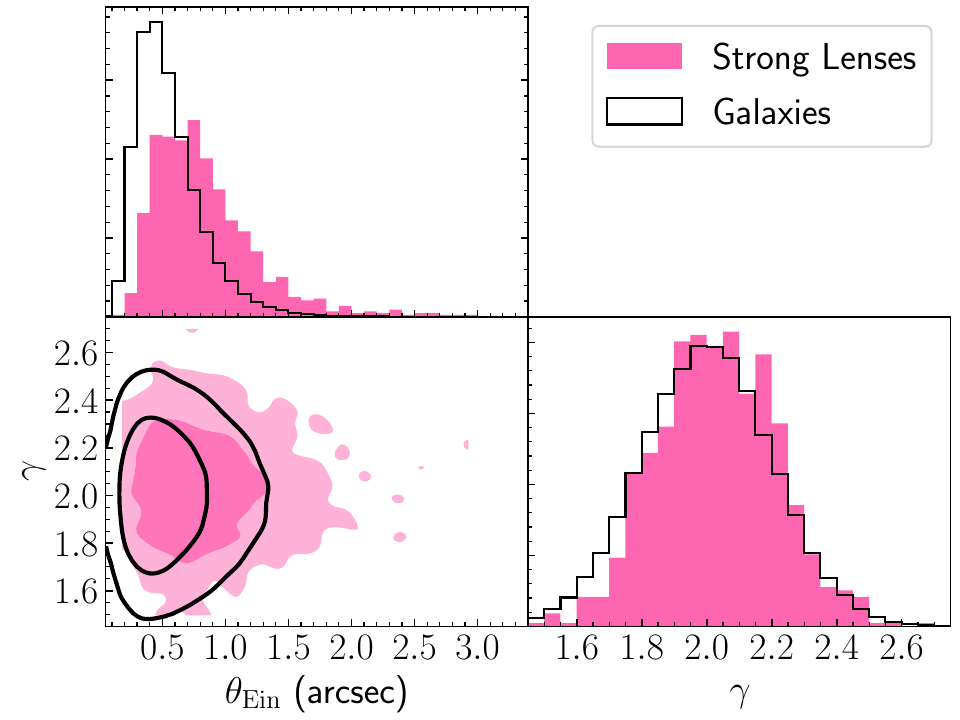}
\caption{
Distribution in Einstein radius and density slope of the strong lens sample (filled contours and histograms) and of the parent population of galaxies (solid lines).
Contour levels enclose 68\% and 95\% of the sample. The histograms are normalised by the number of objects.
\label{fig:teingamma}
}
\end{figure*}

\Fref{fig:teingamma} shows the distribution in Einstein radius and density slope of the lens sample, compared to that of the parent population of galaxies.
The lenses have on average a larger Einstein radius. 
This can be easily understood: at fixed density slope, a galaxy with a larger Einstein radius has a larger lensing cross-section (to be precise, $\crosssect \propto \tein^2$), and therefore is more likely to be a lens.
Additionally, the average value of $\gamma$ of the lens sample is also larger than that of all galaxies (this can be clearly seen in the histogram at the bottom right of \Fref{fig:teingamma}).
This suggests that the lensing cross-section is also an increasing function of $\gamma$.

\begin{figure}
\includegraphics[width=\columnwidth]{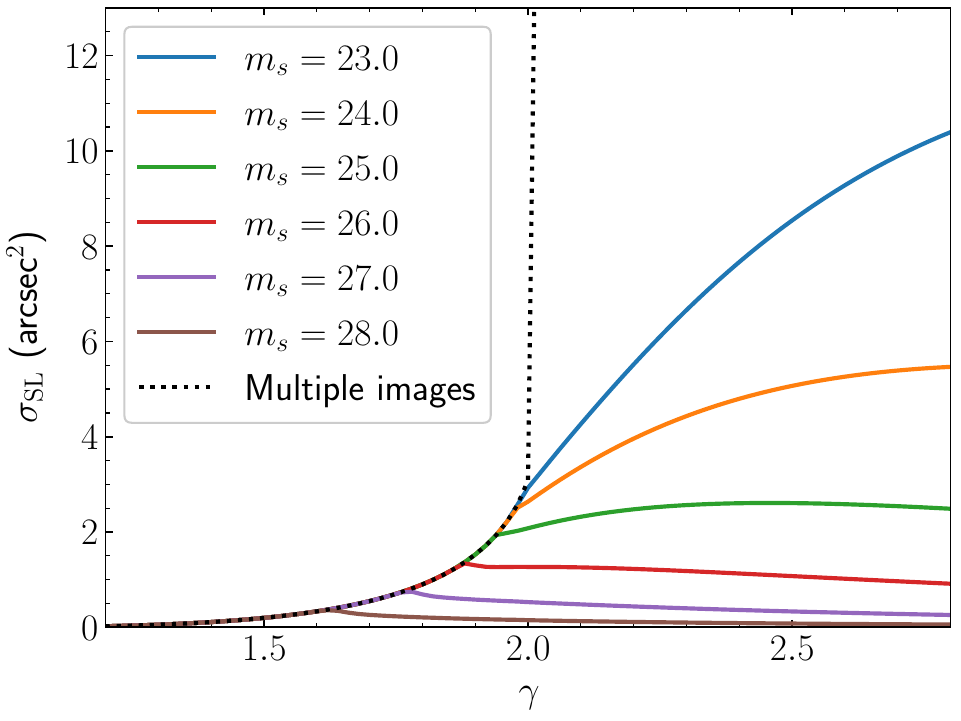}
\caption{
Strong lensing cross-section, defined according to \Eref{eq:crosssect}, as a function of density slope, for a power-law lens with Einstein radius $\tein=1''$.
Coloured lines correspond to different source magnitudes, according to the legend.
The dotted line indicates the area in the source plane that is lensed into multiple images (infinite for $\gamma>2$).
\label{fig:crosssect}
}
\end{figure}

To better illustrate this feature, \Fref{fig:crosssect} shows $\crosssect$ as a function of $\gamma$, for lenses at fixed $\tein=1''$ and sources of different magnitudes.
The dotted line shows the area of the source-plane region within which a source is lensed into multiple images. This can be considered as the lensing cross-section of an infinitely bright source, which is an upper limit on $\crosssect$.

For low values of $\gamma$, $\crosssect$ equals this upper limit, independently of the source magnitude: this means that, for any source position within the radial caustic, the magnification of image 2 is sufficiently large to guarantee its detection.
As $\gamma$ increases, we observe a different behaviour, depending on $m_s$. At faint magnitudes, $\crosssect$ decreases: this means that, with increasing $\gamma$, a larger area of the source plane is mapped into sets of images with a low magnification of image 2. At bright magnitudes, however, image 2 can be detected even when it is de-magnified, and therefore $\crosssect$ increases with $\gamma$.

\begin{figure}
\includegraphics[width=\columnwidth]{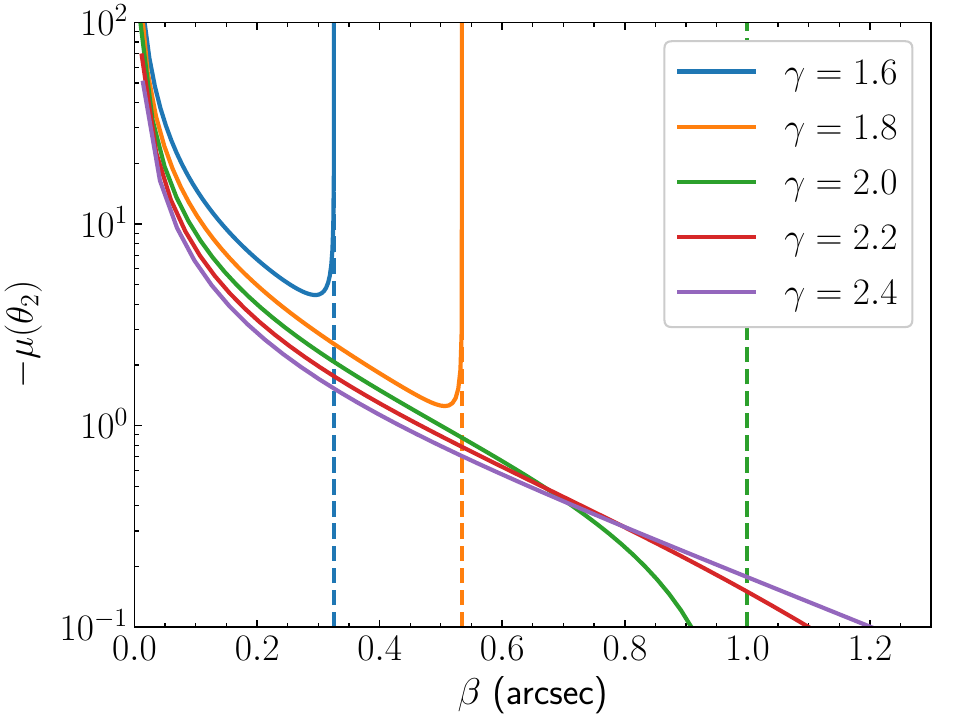}
\caption{
Negative magnification of image 2 as a function of source position, for power-law lenses with $\tein=1''$.
Coloured lines correspond to different values of the density slope $\gamma$, according to the legend.
The dashed lines indicate the radial caustic position for the lenses of the corresponding colour.
\label{fig:muB}
}
\end{figure}

To better understand this behaviour, \Fref{fig:muB} shows the magnification of image 2 as a function of source position, for a few values of $\gamma$.
For $\gamma < 2$, the area of the source plane that produces multiple images (limited by the dashed line) decreases with decreasing $\gamma$, but the magnification of image 2 increases at fixed $\beta$. For $\gamma>2$, the $\mu-\beta$ curves of different lenses cross each other. This crossing is responsible for the different trends of $\crosssect$ with $\gamma$ of sources with different magnitudes.

The net effect, when averaging over the source population, is that galaxies with a higher $\gamma$ have a larger cross-section, and therefore the mean $\gamma$ of the lens population is larger than that of the parent galaxy population.
This, however, is a result of the particular choice of the source magnitude distribution, and is not a general property of power-law lenses.
For example, with a fainter knee of the source magnitude distribution (a higher value of parameter $m_s^*$ in \Eref{eq:msdist}), the difference in the average density slope between the general population of galaxies and the strong lenses becomes smaller.


\section{Model}\label{sect:model}

The simulation produced the following data:
\begin{itemize}
\item $867$ strong lenses, each with measurements of the two image positions, $\toneobs$ and $\ttwoobs$, and related magnitudes, $\moneobs$ and $\mtwoobs$.
\item $99133$ non-detections.
\end{itemize}
I fitted these data with a population of galaxies with power-law density profile, and a population of background sources described by the same magnitude distribution as that used to generate the sample, \Eref{eq:msdist}.
I described each galaxy with the pair of parameters $(\log{\tein},\gamma)$, and assumed that their population distribution is the following product of Gaussians:
\begin{equation}\label{eq:model}
\prlens(\log{\tein},\gamma|\hyperpars) = \mathcal{N}(\mu_{\mathrm{Ein}},\sigma_{\mathrm{Ein}}^2)\mathcal{N}(\mu_{\gamma},\sigma_{\gamma}^2).
\end{equation}
The population parameters of the model are the mean and standard deviations of these two Gaussians:
\begin{equation}
\hyperpars \equiv \{\mu_{\mathrm{Ein}},\sigma_{\mathrm{Ein}},\mu_{\gamma},\sigma_{\gamma}\}.
\end{equation}
The distribution is expressed in terms of the logarithm of the Einstein radius, rather than $\tein$, as this was the form used in section \ref{ssec:lenses} to generate the galaxy population.
In general, a log-normal distribution in $\tein$ is preferred over a simple Gaussian in $\tein$ when the Einstein radius distribution of the sample is not limited to a well-defined order of magnitude.
This is the case, for example, when both galaxy-scale and group- or cluster-scale lenses are included in the population.

This model reproduces exactly the true distribution that was assumed when creating the mock, for $\hyperpars = \{-0.3,0.2,2.0,0.2\}$. In other words, the model is accurate by construction. This makes it possible to test for the accuracy of the inference method of \Sref{sect:theory}.

\begin{figure}
\includegraphics[width=\columnwidth]{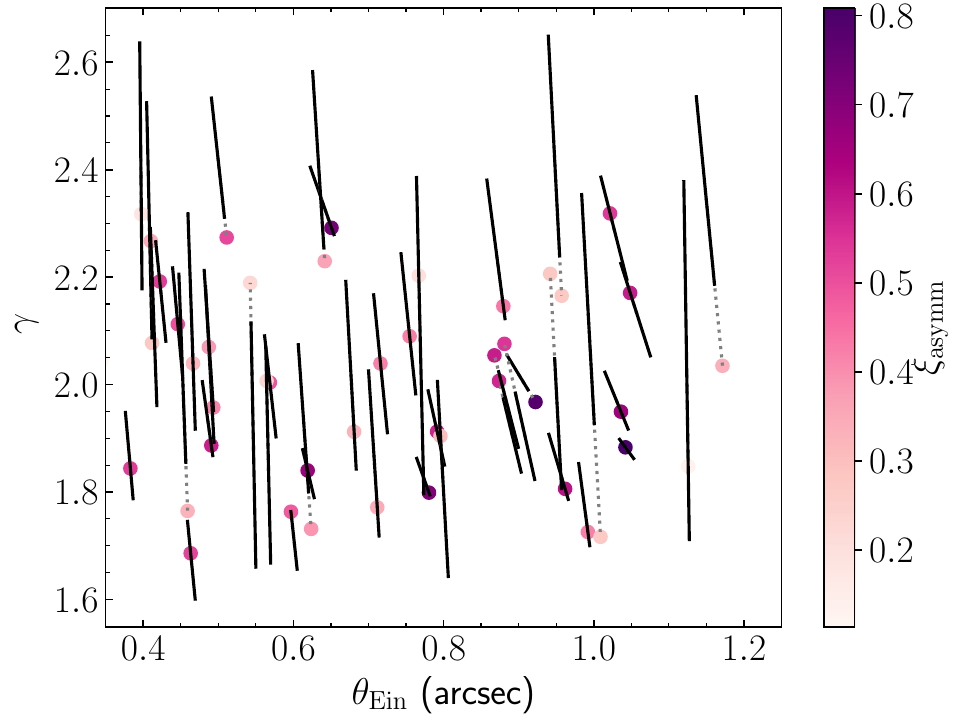}
\caption{
Inference on the Einstein radius and density slope of a subset of lenses, obtained assuming flat priors on all of the lens model parameters.
The black bars indicate the 68\% credible region.
Circles are true values of $\tein$ and $\gamma$ of the lenses, colour-coded by the image asymmetry parameter $\asymm$, defined in \Eref{eq:asymm}.
The grey dotted lines connect the true values to the corresponding measurement. 
\label{fig:individ}
}
\end{figure}

Each lens has four degrees of freedom, $\tein$, $\gamma$, $\beta$ and $m_s$, and four observational constraints: the two image positions and magnitudes.
Therefore, unlike in \citetalias{S+C21}, the model is fully constrained on an individual lens basis. The two magnitudes, however, are only measured with a precision of $0.1$~mag. When fitting a power-law model to an individual lens, this translates into an uncertainty of on the density slope $\gamma$ in the range $0.05-0.3$, depending on the image configuration.
This is illustrated in \Fref{fig:individ}, which shows the 68\% credible region on the Einstein radius and density slope of a subset of lenses, obtained by assuming flat priors on all of the parameters.
The circles in \Fref{fig:individ} show the true values of $\tein$ and $\gamma$ of the lenses, colour-coded by the image asymmetry parameter, which is defined as
\begin{equation}\label{eq:asymm}
\asymm = \frac{\theta_1 + \theta_2}{\theta_1 - \theta_2}.
\end{equation}
Lenses with more asymmetric configurations (darker points) allow for a more precise inference on $\gamma$, though with a stronger degeneracy with $\tein$.
These individual posteriors are just shown for illustrative purposes and are not used in the analysis.

I fitted the model first to the sample of strong lenses, and then to the whole sample, including the non-detections.
I assumed a flat prior on each parameter, over the range reported in \Tref{tab:inference}.
I sampled the posterior probability distribution $\pr(\hyperpars|\{\data\})$ with a Markov Chain Monte Carlo (MCMC) method.

The main technical challenge in this experiment is the computation of the likelihood.
Each system in the sample contributes to the likelihood with a multi-dimensional integral: \Eref{eq:likeint1} (when using only lenses) or \Eref{eq:lensdetint} (when using the whole sample).
In the next two sections, I explain the procedure that I adopted to compute these integrals.

\subsection{Likelihood computation: lenses-only case}\label{ssec:lensint}

When fitting the observed data with the power-law model introduced in this section, \Eref{eq:likeint1} becomes the following 4-dimensional integral,
\begin{align}\label{eq:slint}
\pr(\data|\hyperpars) = A(\hyperpars) \int & d\gamma dm_s d\log{\tein} d\beta\, \pr(\toneobs,\ttwoobs|\tein,\gamma,\beta) \times \nonumber \\ 
& \times \pr(\moneobs,\mtwoobs|\tein,\gamma,\beta,m_s)\pr(\tein,\gamma|\hyperpars) \times \nonumber \\
& \times \pdet(\tein,\gamma,m_s,\beta|S)\pr(m_s),
\end{align}
where $A(\hyperpars)$ is a normalisation constant defined as follows:
\begin{align}\label{eq:normint}
A(\hyperpars)\int d\gamma dm_s d\log{\tein} d\beta\, & \pr(\tein,\gamma|\hyperpars) \pdet(\tein,\gamma,m_s,\beta|S) \times \nonumber \\
& \times \pr(m_s) = 1.
\end{align}

At fixed $\tein$, $\gamma$ and $m_s$, the integral over $\beta$ in \Eref{eq:normint} gives the strong-lensing cross-section defined in \Eref{eq:crosssect}.
The full integral of \Eref{eq:normint} is therefore the population-averaged $\crosssect$.
This integral separates into a part that depends only on $\tein$, and a part that depends on $(\gamma,m_s)$. I computed the former with a general purpose 1D integration method, and the latter via Monte Carlo integration.

Let us now consider the image position likelihood term in \Eref{eq:slint}. Assuming, as in \citetalias{S+C21}, that the image positions are measured exactly, we have
\begin{align}
\pr(\toneobs,\ttwoobs|\tein,\gamma,\beta) = & \delta\left(\theta_1(\tein,\gamma,\beta) - \toneobs\right) \times \nonumber \\
& \delta\left(\theta_2(\tein,\gamma,\beta) - \ttwoobs\right),
\end{align}
where $\theta_1$ and $\theta_2$ are the values of the image positions predicted by the model.
At fixed $\gamma$ and $m_s$, the above product of Dirac delta functions can be easily eliminated after the following variable change,
\begin{equation}\label{eq:change}
(\log{\tein},\beta) \rightarrow (\theta_1,\theta_2),
\end{equation}
and subsequent integration over $\theta_1$ and $\theta_2$.
The result of this operation is
\begin{align}
\pr(\data|\hyperpars) = A(\hyperpars) \int & d\gamma dm_s\, \pr(\moneobs,\mtwoobs|\teinein,\betaein,m_s) \pr(\teinein,\gamma|\hyperpars) \times \nonumber \\
& \pdet(\teinein,\gamma,m_s,\betaein|S)\pr(m_s) \left\lvert\detJ\right\rvert_{\left(\tein,\beta\right)=\left(\teinein,\betaein\right)},
\end{align}
where $\teinein$ and $\betaein$ are the values of the Einstein radius and source position needed to reproduce the observed image positions, and $\detJ$ is the Jacobian determinant of the variable change of \Eref{eq:change}.

At fixed $\gamma$, the integral over $m_s$ is independent of the population parameters $\hyperpars$. I computed this integral for each lens, on a grid of values of $\gamma$, with a spline approximation and integration, prior to running the MCMC.
Finally, I computed the integral over $\gamma$ via spline integration, for each lens at each draw of parameters $\hyperpars$.

\subsection{Likelihood computation: full sample case}

The calculation of the likelihood terms of \Eref{eq:likeprod} takes different avenues, depending on whether the object is a lens or not.
In the case of a non-detection, the likelihood of observing the data given the population parameters is obtained by combining \Eref{eq:likeintnosource}, \Eref{eq:likenonlens} and \Eref{eq:poiss}. The result is
\begin{equation}\label{eq:plnondetlike}
\pr(\data|\hyperpars) = \int d\gamma d\log{\tein} e^{-\lambda} \pr(\tein,\gamma|\hyperpars),
\end{equation}
where $\lambda$ is the expected number of lensed sources given the galaxy parameters.
For a power-law lens, this quantity can be written as
\begin{equation}
\lambda = \nbkg\left(\frac{\tein}{1''}\right)^2\int d m_s \pr(m_s)\crosssect(\tein=1'',\gamma,m_s|S),
\end{equation}
where the lensing cross-section in the integrand is evaluated at a unit Einstein radius.
In other words, the dependence of $\lambda$ on the Einstein radius is analytical.
To obtain $\lambda$ as a function both $\tein$ and $\gamma$, I computed the integral of \Eref{eq:plnondetlike} on a grid of values of $\gamma$ and then interpolated the result with a spline polynomial.
This calculation was done once, before running the MCMC on the population parameters.
I then computed the integral of \Eref{eq:plnondetlike} via Monte Carlo integration.
Since the non-lenses of the simulation are indistinguishable from one another, this integral is the same for all non-detections. Therefore it only needs to be computed once (but with high precision), for each draw of $\hyperpars$.

In the case of a strong lens, the likelihood is given by \Eref{eq:lensdetint}, which for a power-law lens model becomes
\begin{align}
\pr(\data|\hyperpars) = \int & d\gamma d\log{\tein} dm_s d\beta \,\frac{\lambda^{\nsource}e^{-\lambda}}{\nsource!} \pr(\toneobs,\ttwoobs|\tein,\gamma,\beta) \nonumber \\
& \times \pr(\moneobs,\mtwoobs|\tein,\gamma,\beta,m_s)\pr(\tein,\gamma|\hyperpars) \\
& \times \pr(m_s,\beta|{\rm Lens},\tein,\gamma,S).
\end{align}
Using the same method as in section \ref{ssec:lensint}, I reduced this to a two-dimensional integral over $\gamma$ and $m_s$, which I then computed via spline integration.

Out of the $867$ strong lenses, $13$ have $\nsource=2$. For the purpose of evaluating the Poisson likelihood term of \Eref{eq:poiss}, I considered these as double-source lenses. However, I only used image positions and magnitudes for one source, picked randomly.
In a real case scenario, all of the available lensed sources would be used for the analysis, as multiple sources allow for more precise constraints on the lens model parameters. 
That, however, would result in a few lenses having extremely precise inferences of $\gamma$, which would then dominate the inference on the population parameters.
I chose to only use a single source in order to give equal weight to each lens on the inference, which makes the interpretation of the results easier.


\section{Results}\label{sect:results}

\begin{figure*}
\includegraphics[width=\textwidth]{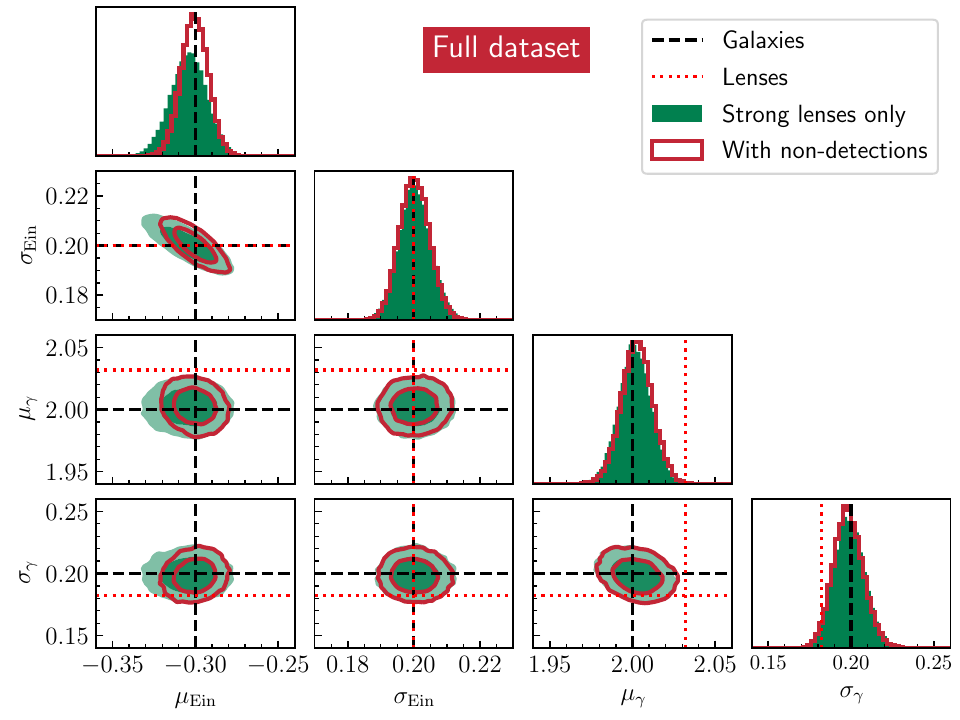}
\caption{
Posterior probability distribution of the galaxy population model parameters, $\hyperpars$, obtained by fitting only the strong lenses (green filled contours) and the full sample (red solid contour lines). All of the available lensing data (image positions and magnitudes) were used for the fit.
Black dashed lines mark the true values of the parameters.
Red dotted lines indicate the values of the parameters obtained by fitting the model of \Eref{eq:model} to the subsample of strong lenses (the mean $\log{\tein}$ parameter value is $\mu_{\mathrm{Ein}} = -0.12$, which is outside of the plot range).
\label{fig:cp1}
}
\end{figure*}

\Fref{fig:cp1} shows the posterior probability distribution of the population model parameters $\hyperpars$, obtained by fitting the strong lenses only (green filled contours) and the entire sample of galaxies (red solid lines).
The median and 68\% credible region of the marginal posterior probability of each parameter is reported in \Tref{tab:inference}.
In both cases, the model is able to recover the true distribution of $\tein$ and $\gamma$ of the galaxy sample (dashed black lines).

\Fref{fig:cp1} also shows the values of the model parameters of the population of lenses (dotted red lines), which I obtained by fitting the model of \Eref{eq:model} directly to the true values of $\tein$ and $\gamma$ of the $867$ lenses.
The average value of $\gamma$ of the lens population is $2.03$.
This is the value of $\mu_{\gamma}$ that one would infer by fitting the lens sample without taking into account selection effects \citepalias[that is, following the method of][]{S+C21}.
Thanks to the formalism introduced in this paper, this solution is ruled out at the $3\sigma$ level, and the true value $\mu_{\gamma}=2.0$ is recovered.
\begin{table*}
\caption{Inference on the model parameters.
}
\label{tab:inference}
\begin{tabular}{ccccccccl}
\hline
\hline
Parameter & Truth & Lenses & Prior & \multicolumn{2}{c}{Fit to full dataset} & \multicolumn{2}{c}{Fit to image positions only} & Description \\
 & & & & Lenses only & Full sample & Lenses only & Full sample & \\
\hline
$\mu_{\mathrm{Ein}}$ & $-0.30$ & $-0.12$ & $U(-0.5,0.5)$ & $-0.297_{-0.010}^{+0.010}$ & $-0.298_{-0.008}^{+0.007}$ & $-0.298_{-0.011}^{+0.010}$ & $-0.295_{-0.011}^{+0.011}$ & Mean $\log{\tein}$ \\
$\sigma_{\mathrm{Ein}}$ & $0.20$ & $0.20$ & $U(0.0,0.3)$ & $0.199_{-0.004}^{+0.004}$ & $0.201_{-0.004}^{+0.004}$ & $0.200_{-0.004}^{+0.004}$ & $0.201_{-0.004}^{+0.004}$ & Scatter in $\log{\tein}$ \\
$\mu_{\gamma}$ & $2.00$ & $2.04$ & $U(1.8,2.2)$ & $1.997_{-0.009}^{+0.009}$ & $2.000_{-0.009}^{+0.009}$ & $1.97_{-0.10}^{+0.09}$ & $1.97_{-0.04}^{+0.04}$ & Mean $\gamma$ \\
$\sigma_{\gamma}$ & $0.20$ & $0.18$ & $U(0.0,0.5)$ & $0.201_{-0.008}^{+0.008}$ & $0.201_{-0.008}^{+0.008}$ & $0.20_{-0.07}^{+0.06}$ & $0.19_{-0.03}^{+0.03}$ & Scatter in $\gamma$ \\

\end{tabular}
\tablefoot{
Column (2): true values of the population parameters. 
Column (3): values of the lens population parameters, obtained by fitting the model of \Eref{eq:model} to the sample of detected strong lenses.
Column (4): priors on the parameters.
Column (5): median, 16th and 84th percentile of the marginal posterior probability distribution of each parameter obtained by fitting the sample of strong lenses with the model of \Sref{sect:model}.
Column (6): inference on the full sample (lenses and non-detections).
Column (7): inference using image positions only (no magnitudes), lens-only sample.
Column (8): inference using image positions only, full sample.
}
\end{table*}
\begin{figure*}
\includegraphics[width=\textwidth]{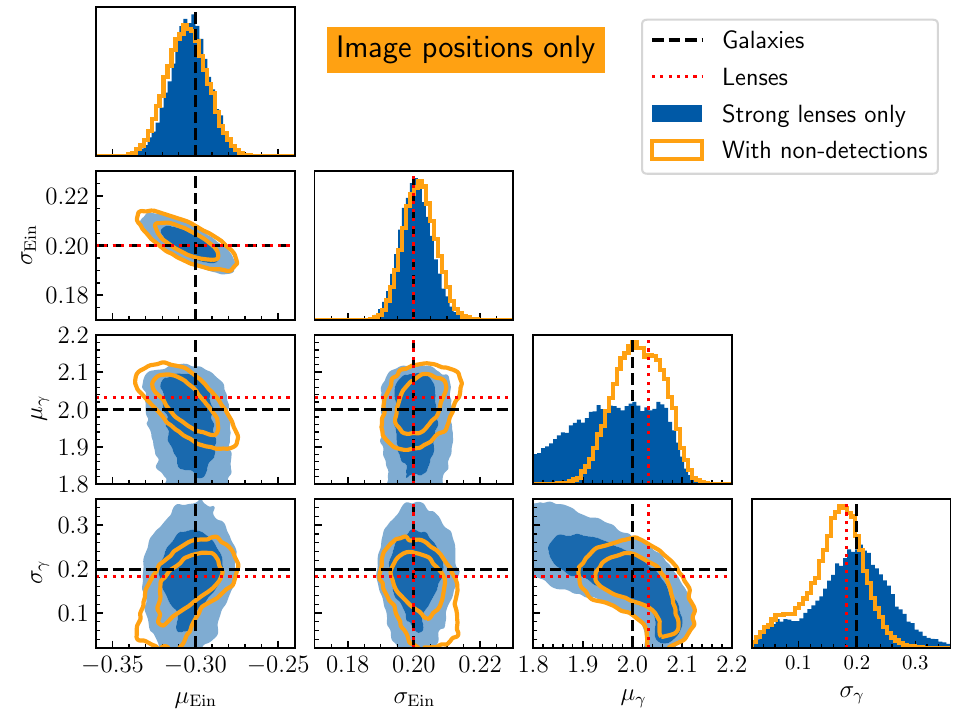}
\caption{
Same as \Fref{fig:cp1}, but obtained by fitting only the image positions (no magnitudes) of the lenses.
\label{fig:cp2}
}
\end{figure*}

The posterior probability obtained when using the full galaxy sample (red solid lines in \Fref{fig:cp1}) is almost identical to the one based purely on the sample of lenses.
The full sample brings in additional information: the number of non-detections and the number density of background sources, $\nbkg$.
This result, however, shows that this information contributes little to the posterior, which is dominated by the strong lenses, with the exception of the parameter $\mu_{\mathrm{Ein}}$.

The statement above is conditional on the type and quality of the data that the model is fitted to.
For the analysis presented so far, I assumed that, for each lens, image positions and magnitudes are available. The magnitudes of the two images provide a constraint on the ratio of the magnifications at positions $\theta_1$ and $\theta_2$, which in turn can be used to constrain the density slope \citep[see][]{ORi++19,ORi++20}.
Often, however, statistical strong lensing analyses do not make use of magnification information \citep[see for example][]{Aug++10,Bol++12,Son++15,Son++19}, which usually requires high-quality imaging data to be extracted.

I investigated an alternative scenario, in which the only lensing data available are the image positions and the knowledge of the lens detection probability, $\pdet$, as a function of the model parameters.
Since the selection criterion of the simulated sample, \Eref{eq:maxmag}, relies on the magnitude of the second image, it is difficult to imagine a situation in which $\pdet$ is known but the image magnitudes are not. 
However, the rationale of this experiment is to simulate a more general scenario in which the selection function is known, but no magnification information is available.

\Fref{fig:cp2} shows the posterior probability distribution obtained by fitting the same model of \Sref{sect:model} to lensing data, while removing the term related to the observed magnitudes, $\pr(\moneobs,\mtwoobs|\tein,\gamma,\beta,m_s)$, from the likelihood.
The inference on the parameters describing the $\gamma$ distribution is much less precise, as a result of the lack of direct constraints on the magnification ratios. 
When using only the strong lenses, I obtained $\mu_{\gamma}=1.97\pm0.10$, which is an order of magnitude more uncertain than what the full dataset allows.
In this case, the addition of the non-detections reduces this uncertainty by a factor of two: $\mu_{\gamma} = 2.01\pm0.05$.

Interestingly, image position data are still able to constrain $\mu_{\gamma}$, to some extent, even when non-detections are not considered.
This is because certain image configurations can only be produced by lenses with certain values of $\gamma$: for example, in lenses with $\gamma>2$, the position of image 2 can be arbitrarily close to the lens centre. If $\gamma<2$, however, there is a lower limit to the value of $\theta_2$, set by the position of the radial critical curve.
This explains the shape of the $\mu_\gamma-\sigma_\gamma$ degeneracy observed in the fit to the lens sample (see the panel in the third column, bottom row of \Fref{fig:cp2}): if the average $\gamma$ is smaller than $2$, the scatter in $\gamma$ must be large, in order to allow for the presence of lenses with $\gamma>2$ in the sample, which are needed to reproduce highly asymmetric image configurations.


\section{Discussion}\label{sect:discuss}

\subsection{Advantages of a complete lens sample}

I introduced a formalism for the analysis of strong lensing data from a survey with known selection function, then applied it to a simulated sample of lenses that is complete above a well-defined observational cut.
Sample completeness is not a strict requirement in order to apply the method of \Sref{sect:theory}: in principle, it is sufficient to know the lens detection probability $\pdet$ of the survey.
However, if a fraction of lenses is missed by the survey, in a way that correlates with some lens property and that is not taken into account by the model for $\pdet$, biases will be introduced.

For example, let us suppose that the lens detection efficiency of a survey like the one considered in \Sref{sect:sims} depends not only on the magnitude of image 2, but also on the surface brightness contrast between the lens and the source light.
Let us suppose that lenses with a small value of $\theta_2$ are, for this reason, missed by the survey. If this selection is not taken into account, the model might be biased against solutions with large values of $\gamma$, which are instead needed to reproduce lenses with highly asymmetric configurations.
Having a complete lens sample guarantees against such a possibility.
In this example case, this requires ensuring that no lenses are missed, regardless of their image configuration.

\subsection{Technical challenges}

In order to apply this method to real samples of lenses, several challenges need to be overcome.
The most important one is obtaining an accurate description of the lens detection probability, in order to avoid biases like the one in the example above.
The experiment that I carried out assumed point-like sources, but virtually all strongly-lensed sources have non-zero size. Lensed galaxies are usually well-resolved, which means that their detection depends on their surface brightness distribution.
The source parameter set, $\psisource$, will then need to include quantities such as the half-light radius and the S\'{e}rsic index, in addition to the magnitude. 

As new parameters are introduced, the model describing the source population, $\pr(\psisource)$, must be updated accordingly. While, in principle, $\psisource$ can be measured on the large population of galaxies that are not strongly lensed, in practice the lens detection probability could depend on features that are below the resolution limit of non-lensed galaxies. For example, the presence of bright clumps can boost the detectability of a lens, but the appearance of a clumpy galaxy changes when it is strongly lensed, due to the magnification effect \citep[see for example][]{Fau++21}.
It is possible, in principle, to generalise the model of \Sref{sect:model} and infer the source property distribution $\pr(\psisource)$ simultaneously with the lens distribution. In practice, this would increase the computational burden of the inference procedure. 

In case of a lens sample selected on the basis of the availability of spectroscopic redshifts, then the selection function needs to also take into account the efficiency of obtaining such measurements.
On the one hand, this can complicate the selection function even further, as in the case of the SL2S survey discussed in \Sref{sect:intro}.
On the other hand, if the primary means of the lens search is spectroscopic, meaning that the presence of a lensed source is assessed with the same observational data used to measure its redshift, then the image detection and source redshift measurement aspects of the problem can be merged.

This is the case, for example, of the Sloan ACS Lens Survey \citep[SLACS,][]{Bol++06}, and the SINFONI Nearby Elliptical Lens Locator Survey \citep[SNELLS,][]{Smi++15}.
\citet{Arn++12} carried out a study of the selection function of SLACS, which, if extended, could be used to carry out an analysis like the one presented in this paper.
SNELLS, and the more recent MNELLS \citep{Col++20}, are particularly suited for the application of this method, in virtue of their clean selection function.

One class of systems for which it is easier to obtain the selection function, compared to lensed galaxies, is that of strongly lensed quasars.
Current lensed quasar searches are already able to produce highly complete lens samples \citep[see for example][]{A+S19}
It could be possible, in the near future, to identify a complete set of strongly-lensed quasars: for instance, by carrying out a systematic search on a well-defined sample of massive galaxies, applying a selection criterion such as that of \Eref{eq:maxmag}, and obtaining spectroscopic follow-up observations to obtain redshifts and rule out contaminants \citep[as was done by][]{Agn++18}.

Lensed quasars, however, are subject to microlensing, which introduces a source of scatter in the relation between the intrinsic source magnitude and the observed one.
Microlensing depends on the fraction of mass in compact objects (that is, stars) and on the size of the light-emitting region of the quasar, which is the accretion disk.
A rigorous analysis would require modelling microlensing, in order to obtain an accurate selection function of lensed quasars.
This would also provide an opportunity to further constrain the stellar mass-to-light ratio of the lens galaxies \citep{Ver19}. There is, however, an associated cost in terms of model complexity.



If the lens detection probability is a function of image configuration, whether directly, or, as in the simulation of \Sref{sect:sims}, indirectly (highly asymmetric configuration are rarely detected, because of the de-magnification of image 2), then the lens population model, $\pr(\psilens)$, needs to be extended.
This is because, unlike in the experiment carried out in this work, real lenses are not axisymmetric, and the ellipticity of a lens can change the number, the shape and the magnification of the strongly-lensed images.
At the very least, it is necessary to add two degrees of freedom to the model: the axis ratio and orientation of the mass component.
Moreover, lens models often require to account for external shear in order to correctly reproduce the observed images. Galaxy ellipticity and external shear could be incorporated in the hierarchical model used to describe the lens population, but this would make computations more challenging (especially the evaluation of integrals of the kind of \Eref{eq:likeint1}). 

In summary, most practical applications of the method proposed in this paper will likely require an extension of the model considered so far.
This will pose important computational challenges, analogous to those discussed in section 6.5 of \citetalias{S+C21}.

\subsection{Other applications}

In addition to providing unbiased descriptions of the structural parameter distribution of galaxies, this method can be used for combining information from different samples of lenses.
One important case in which this is needed is the analysis of time-delay lenses for the study of cosmology. 
The current approach of the leading team in this field, the TDCOSMO\footnote{\url{http://tdcosmo.org}} collaboration, is that of using information obtained from samples of strong lenses with no time delays as a prior on a smaller set of time-delay lenses \citep{Bir++20}. In Paper II \citep{Son21}, I showed how, with such an approach, it is possible to obtain $1\%$-level measurements of the Hubble constant.

\Eref{eq:one}, however, shows that two different lens samples, $\prslone$ and $\prsltwo$, with different source populations, $\prsourceone$ and $\prsourcetwo$, and selection functions $\pdetone$, $\pdettwo$, are biased in different ways with respect to the general population of galaxies $\prlens$.
Therefore, using information relative to one lens sample as a prior on the other might lead to biases (the amplitude of which needs to be quantified with dedicated simulations).

The formalism introduced in this work offers a solution to this problem: if the parent population of galaxies, among which strong lenses are searched, is selected in the same way in the two surveys, then $\prlens$ is by construction the same distribution for the two samples. 
It is relatively easy to achieve this: usually, galaxy populations are selected by means of cuts applied at the catalogue level (for example, by imposing that galaxies lie in a given redshift range and are brighter than a luminosity threshold), so it is sufficient to apply the same cuts consistently to the two samples.
In that case, it is possible, with the method of \Sref{sect:theory}, to infer $\prlens$ from one sample and then use it as a prior on the other sample.

Another possible application is the analysis of strong lenses with no spectroscopic measurement of the source redshift. Although for the experiment carried out in this work I assumed that the source redshifts were known, this is not a strict requirement. 
The main requirement is for the source parameter distribution, $\pr(\psisource)$, which includes the source redshift, to be known. 
If that is the case, then the source redshift can be treated as a nuisance parameter and marginalised over. This is the subject of Paper IV \citep{Son22}.

Finally, information from the relative number of lenses and non-detections can in principle be used to break the mass-sheet degeneracy.
This is because, in general, a mass-sheet transformation \citepalias[see section 2.1 of][ for a definition]{Son21} changes the strong lensing cross-section of a galaxy.
I tried to explore this possibility on the sample generated in \Sref{sect:sims}.
The change in $\crosssect$ with the mass-sheet transformation parameter, however, varies with source magnitude, and any signal is washed out after averaging over the source population.

\section{Conclusions}\label{sect:concl}

This paper introduced the formalism necessary for inferring the structural properties of a sample of galaxies from the observations of a subset of strong lenses drawn from it. 
Although the strong lenses are a biased sample, when the properties of the background source population and the selection function of the sample are known, it is possible, with this method, to remove this bias.

There are two variations of the method: one that uses only the strong lenses, and one that combines the lens sample with non-detections.
I tested both of these methods on a simulated sample of galaxies that included a complete set of lenses. 
In both cases, the true properties of the general galaxy population were accurately recovered.

The experiment also revealed that, when the lensing data provides constraints on the relative magnification between the images, in addition to the image positions, then the non-detections add little constraining power.
Nevertheless, in such cases, the number of non-detections could be used as a consistency check, to test the model assumptions.
However, when magnification constraints are not available, non-detections reduce model uncertainties significantly.

There are several technical challenges to be overcome, in order to apply this method to real samples of lenses.
Nevertheless, the formalism developed in this work constitutes the first necessary step in order to use strong lenses to make unbiased statements on the galaxy population.




\bibliographystyle{aa}
\bibliography{references}

\end{document}